\documentclass[journal=nalefd,manuscript=letter,layout=twocolumn,email=false]{achemso}
\pdfoutput=1

\usepackage{siunitx}
\usepackage{amsmath}
\usepackage[utf8]{inputenc}

\let\oldmaketitle\maketitle
\let\maketitle\relax

\author{Mátyás Kocsis}%
\affiliation{Department of Physics, Budapest University of Technology and Economics and MTA-BME Lendület Nanoelectronics Research Group, Budafoki út 8, 1111 Budapest, Hungary}%
\author{Oleksandr Zheliuk}%
\affiliation{Zernike Institute for Advanced Materials, University of Groningen, Nijenborgh 4, 9747 AG Groningen, the Netherlands}%
\author{Péter Makk}%
\affiliation{Department of Physics, Budapest University of Technology and Economics and MTA-BME Lendület Nanoelectronics Research Group, Budafoki út 8, 1111 Budapest, Hungary}%
\author{Endre Tóvári}%
\affiliation{Department of Physics, Budapest University of Technology and Economics and MTA-BME Lendület Nanoelectronics Research Group, Budafoki út 8, 1111 Budapest, Hungary}%

\author{Péter Kun}%
\affiliation{Institute of Technical Physics and Materials Science, MFA, Centre for Energy Research, Hungarian Academy of Sciences, P.O. Box 49, 1525 Budapest, Hungary}%

\author{Oleg Evgenevich Tereshchenko}%
\affiliation{St. Petersburg State University, 198504, St. Petersburg, Russia.}%
\alsoaffiliation{A.V. Rzhanov Institute of Semiconductor Physics, 630090, Novosibirsk, Russia.}%
\alsoaffiliation{Novosibirsk State University, 630090, Novosibirsk, Russia.}%
\author{Konstantin Aleksandrovich Kokh}%
\affiliation{St. Petersburg State University, 198504, St. Petersburg, Russia.}%
\alsoaffiliation{Novosibirsk State University, 630090, Novosibirsk, Russia.}%
\alsoaffiliation{V. S. Sobolev Institute of Geology and Mineralogy, 630090, Novosibirsk, Russia.}%

\author{Takashi Taniguchi}%
\affiliation{International Center for Materials Nanoarchitectonics, National Institute for Materials Science,  1-1 Namiki, Tsukuba 305-0044, Japan}%
\author{Kenji Watanabe}%
\affiliation{Research Center for Functional Materials, National Institute for Materials Science, 1-1 Namiki, Tsukuba 305-0044, Japan}%

\author{Justin Ye}%
\affiliation{Zernike Institute for Advanced Materials, University of Groningen, Nijenborgh 4, 9747 AG Groningen, the Netherlands}%
\author{Szabolcs Csonka}%
\affiliation{Department of Physics, Budapest University of Technology and Economics and MTA-BME Lendület Nanoelectronics Research Group, Budafoki út 8, 1111 Budapest, Hungary}%

\title[]{In situ tuning of symmetry-breaking induced non-reciprocity in giant-Rashba semiconductor BiTeBr}%

\begin{document}

\twocolumn[
\begin{@twocolumnfalse}
\oldmaketitle
\begin{abstract}
Non-reciprocal transport, where the left to right flowing current differs from the right to left flowing one, is an unexpected phenomenon in bulk crystals.
BiTeBr is a non-centrosymmetric material, with a giant Rashba spin-orbit coupling which presents this unusual effect when placed in an in-plane magnetic field.
It has been shown that this effect depends strongly on the carrier density, however, in-situ tuning has not yet been demonstrated.
We developed a method where thin BiTeBr flakes are gate tuned via ionic-liquid gating through a thin protective hBN layer.
Tuning the carrier density allows a more than \SI{400}{\percent} variation of the non-reciprocal response.
Our study serves as a milestone on how a few-atomic-layer-thin van der Waals protection layer allows ionic gating of chemically sensitive, exotic nanocrystals.
\end{abstract}
\end{@twocolumnfalse}
]

Tuning the carrier density of nanostructures is essential for various applications
and the exploration of exotic scientific phenomena.
It lies in the heart of the operation of field effect transistors,
it allows the implementation of electron optical elements in graphene,\cite{taychatanapat_electrically_2013, grushina_ballistic_2013, rickhaus_ballistic_2013, rickhaus_guiding_2015}
and affects valley excitons in 2D materials,\cite{wang_colloquium_2018}
spin relaxation,\cite{dettwiler_stretchable_2017}
or the exchange coupling in spintronic devices.\cite{ohno_electric-field_2000}
Tuning of the carrier density is typically  achieved by  gate electrodes,
which are separated from the nanostructure by an insulating layer.
Ionic-liquid (IL) gating presents a much more effective alternative,
by inducing a layer of charged ions at the surface of the sample.\cite{bisri_endeavor_2017, petach_disorder_2017, hu_self-organization_2017}
Large gating efficiency of IL enables e.g. two-dimensional Ising superconductivity in semiconductor MoS\textsubscript{2}\cite{lu_evidence_2015, chen_continuous_2018, zheliuk_josephson_2019}
or promotes ferromagnetism in platinum\cite{liang_inducing_2018}.

\begin{figure*}[h!]
	\includegraphics[width=\textwidth]{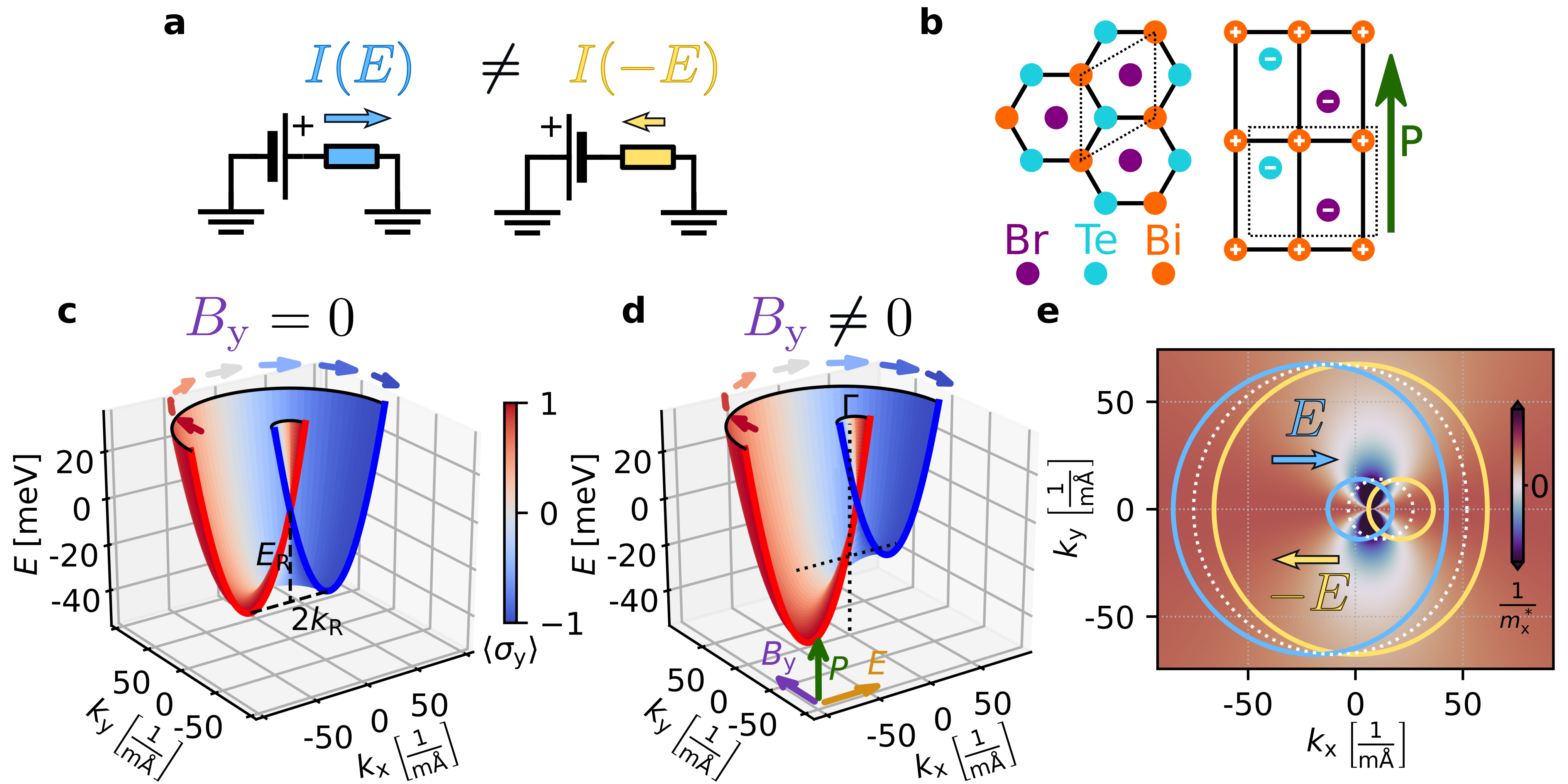}
	\caption[Caption]{
		\emph{a)}~Due to the lack of inversion symmetry of the crystal and the presence of an in-plane magnetic field,
		the current flowing through the sample depends on the polarity of the electric field applied.\cite{ideue_bulk_2017, tokura_nonreciprocal_2018}
		\emph{b)}~The crystal structure of BiTeBr (projected on the 001 and 1120 planes).\cite{shevelkov_crystal_1995, eremeev_rashba_2013, fiedler_termination-dependent_2015}
		The broken inversion symmetry and large polarization ($P$) lead to the gigantic Rashba SOI.\cite{molenkamp_rashba_2001}
		The dashed lines show the border of a possible unit cell.
		\emph{c)}~The band structure of BiTeBr.
		The Rashba energy ($E_\mathrm{R}$) and wavenumber offset ($k_\mathrm{R}$) are indicated,
		their fraction yields $\alpha_\mathrm{R}=\frac{2E_\mathrm{R}}{k_\mathrm{R}}=\SI{2}{\eV\angstrom}$ as the Rashba parameter,\cite{moreschini_bulk_2015, sakano_strongly_2013}
		one of the largest in any bulk material.\cite{ishizaka_giant_2011}
		The strong Rashba SOI causes a helical spin structure to develop,\cite{eremko_spin_2014, maas_spin-texture_2016, ishizaka_giant_2011}
		as indicated by the arrows.
		The band structures shown here assume a two dimensional crystal,
		however the theoretical calculations concern both two and three dimensional cases.\cite{ideue_bulk_2017}
		\emph{d)}~The band structure in an in-plane magnetic field.
		The spins parallel to the magnetic field are shifted to lower energies,
		while the anti-parallel spins are shifted up.
		The direction of the magnetic field $B_y$, electric field $E_x$, and polarization $P$ are indicated.
		\emph{e)}~Fermi surfaces near the energy minimum of the antiparallel spins,
		and colormap of the reciprocal of the effective mass $m^*$.
		Compared to the equilibrium (white dashed line),
		applying an electric field shifts the Fermi surface (indicated by blue and yellow curves).
		Due to asymmetry introduced by the magnetic field and SOI,
		the electrons' effective mass (along the $x$ direction) on each surface is different.
	}
	\label{fig:disp}
\end{figure*}

However, the applicability of IL gating is limited,
as it could induce changes in some systems.\cite{zhang_fabrication_2016, zhang_porous_2016}
In this Letter we show how IL gating combined with a van der Waals protective layer can be applied to chemically sensitive crystals.
In particular, we will demonstrate how the exotic, non-reciprocal resistance of BiTeBr crystals can be boosted by IL gating.
First a brief introduction to the non-reciprocal resistance of BiTeBr is given,
followed by the description of novel devices making use of a few-atomic-layer-thin hBN protection layer,
and IL gating experimental results showing that the non-reciprocal resistance can be tuned by over \SI{400}{\percent}.

In crystals without inversion symmetry, such as BiTeBr, a very surprising non-reciprocity has been observed
in the presence of an in-plane magnetic field ($B$)\cite{ideue_bulk_2017}:
the resistance of the sample depends on the polarity of the applied voltage, as illustrated in Fig.~\ref{fig:disp}.a.
The source of the non-reciprocal behaviour lies in the crystal structure of BiTeBr,\cite{tokura_nonreciprocal_2018} shown in Fig.~\ref{fig:disp}.b.
The alternating layers of (BiTe)\textsuperscript{+} and Br\textsuperscript{-} break inversion symmetry,\cite{fiedler_termination-dependent_2015}
and lead to one of the strongest, so-called giant-Rashba spin-orbit interaction (SOI) in any bulk material\cite{ishizaka_giant_2011}.
Applying an in-plane magnetic field distorts the dispersion relation of the conduction band, as shown in Fig.~\ref{fig:disp}.c and d.
The application of an in-plane electric field ($E$ or $-E$) shifts the distorted Fermi surface,
which will encompass states with different effective masses for the two polarities,
as shown in Fig.~\ref{fig:disp}.e.
As a result, the magnitude of the resulting charge current ($I$) depends on the direction of $E$ with respect to $B$.
Specifically, it depends on the vector product,\cite{tokura_nonreciprocal_2018}
and the voltage-current characteristic takes the form
\begin{equation}
	V = IR_\mathrm{0}\left( 1 + \gamma \mathbf P \cdot \mathbf B \times \mathbf I \right),
	\label{eq:V}
\end{equation}
where $I$ and $B$ are in-plane, $P$ is the polarization of the crystal and is always out-of-plane, $R_\mathrm{0}$ is the resistance at $B=0$,
and $\gamma$ measures the strength of the non-reciprocity.
$\gamma$ is related to the carrier density $n$ by $\gamma\propto\frac{1}{n^2}$,\cite{ideue_bulk_2017}
enabling us to tune non-reciprocity by modifying the carrier density.
We note that based on Equation \eqref{eq:V}, in this realtion $I$ is the asbolute, not RMS amplitude of the AC current.

%%% Samples

\begin{figure}[h!]
	\includegraphics{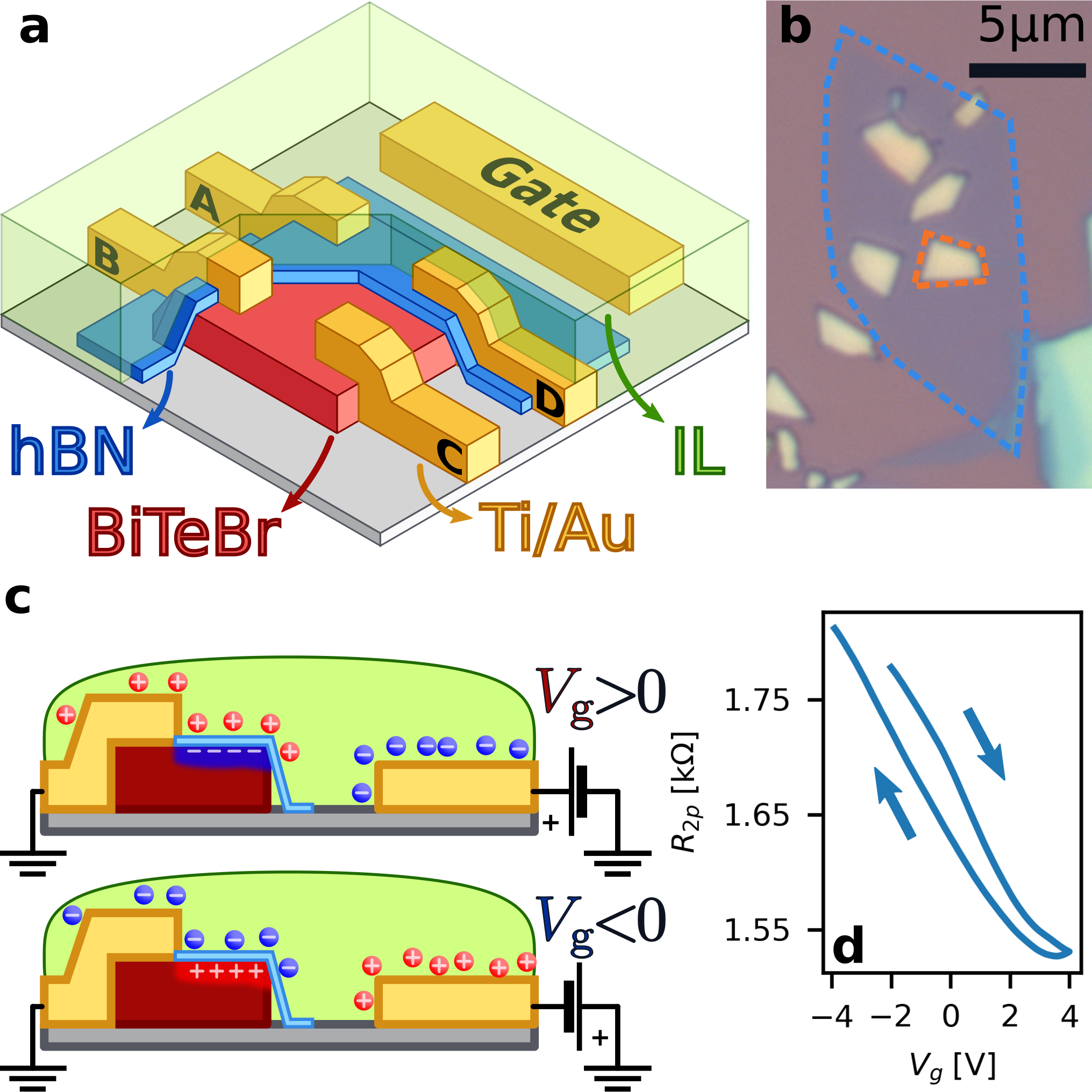}
	\caption{
		\label{fig:sample}
		\emph{a)}~Schematic image of the sample, with the IL and hBN peeled away from the lower corner to show the BiTeBr flake.
		\emph{b)}~Optical image of the BiTeBr samples, after covering it with hBN (blue dashed line).
		\emph{c)}~Schematic image of liquid gating.
		Due to the metallic nature of the sample the bulk is screened,
		only a thin surface layer is affected by the gating.
		\emph{d)}~Two-point resistance as a function of the gate voltage at \SI{220}{\K}.
		The hysteretic behaviour is due to the slow motion of the ions.
	}
\end{figure}

Due to the metallic nature and large carrier density of BiTeBr (\SIrange[]{3e18}{1.3e19}{\per\cm\cubed}),
we have chosen IL gating to control the carrier density.
For ionic-liquid gating the sample and a separate gate electrode are covered by the IL, as shown in Fig.~\ref{fig:sample}.a.
A finite potential is applied to the gate electrode while the sample is grounded.
This generates an accumulation of ions at the sample surface, as shown in Fig.~\ref{fig:sample}.c.
However, we found that BiTeBr is very sensitive to the ions of the IL,
despite using DEME-TFSI as a gating medium, for its chemical stability.
As soon as a nonzero voltage ($V_\mathrm{g}$) was applied to the gate electrode, a current started to flow,
and the BiTeBr flakes decomposed,
as demonstrated in supplementary Fig.~S1.c and d.

To protect the BiTeBr flakes from the IL,
we developed a novel heterostructure where the crystal is protected by a few atomic layer thin hBN flake,
as shown in Fig.~\ref{fig:sample}.a and b.
The measurements presented here were carried out on the flake outlined in orange.
As hBN is chemically stable, this technique can be used with a wide range of ILs,
enabling previously incompatible materials to be used with liquid-gates.
By distancing the ions from the surface the gating also becomes more homogeneous.\cite{petach_disorder_2017}

We chose hBN flakes that were thin enough to allow for effective gating,
but mechanically stable enough to cover the \SIrange{40}{50}{\nm} thick BiTeBr flakes.
For our purpose \SIrange{3}{5}{\nm} thin flakes were selected.\cite{gorbachev_hunting_2011}
Fig.~\ref{fig:sample}.b shows an assembled stack, with the hBN flake outlined in blue and the measured BiTeBr flake in orange.
BiTeBr flake composition was confirmed by Raman and EDS analysis, to sort out frequently occuring impurities, as discussed in Methods and the Supplementary Information.

Ti/Au contacts were used to contact the BiTeBr flake.
Before evaporating the metallic contacts, the hBN underneath was etched away.
The same PMMA mask was used for the etching and the evaporation.
This insures that the whole BiTeBr flake is covered either by the electrodes or hBN,
and no ions can leak in at the hBN-electrode interface.
Device fabrication is detailed in \textit{Methods}.

The two-point resistance of BiTeBr as a function of the gate voltage ($V_\mathrm{g}$) is shown in Fig.~\ref{fig:sample}.d.
The measurement was carried out at \SI{220}{\K}, above the glass transition temperature of DEME-TFSI.
The change in the resistance shows that the gating is successful.
The leakage current through the gate electrode was continuously monitored throughout the measurement,
and never exceeded \SI{0.3}{\nA} as is shown in Fig.~S1.b of the Supporting Information,
confirming the isolation of BiTeBr from the IL by the hBN layer.

%%% 2nd harm. meas.

\begin{figure}[h!]
	\includegraphics{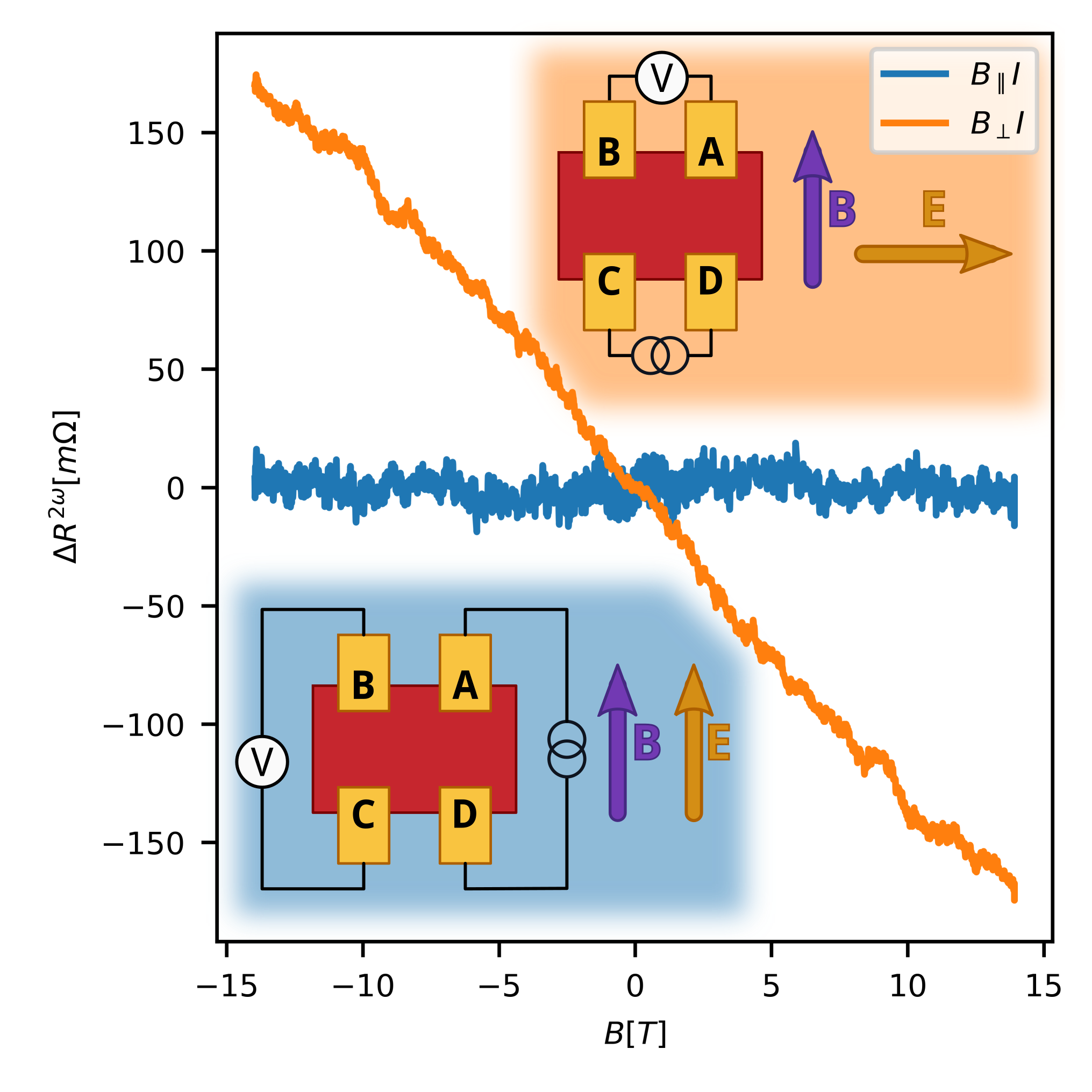}
	\caption{
		\label{fig:nonil2ndharm}
		Second harmonic measurements before the IL gate is applied,
		showing the anti-symmetrized second-harmonic signal.
		The relative orientation of the magnetic and electric field is shown in the insets.
		$\gamma$ can be calculated from the slope of the second harmonic resistance using Equation \eqref{eq:V}.
		As expected, no signal is detected in the parallel case.
	}
\end{figure}

Now we will turn to the measurement of the non-reciprocity.
As shown in Equation \eqref{eq:V}, the non-reciprocity leads to the non-linear current voltage characteristics.
This non-linearity can be easily observed by applying an AC current and measuring the second harmonic response.
While other effects such as magnetoresistance,\cite{tokura_nonreciprocal_2018}
or non-trivial thermoelectric behaviour\cite{nakai_nonreciprocal_2019} due to heating at the contacts,
can generate second harmonic signals of their own,
all those scale with even powers of $B$,
and can be avoided by taking only the antisymmetric part of the second harmonic resistance,
$\Delta R^{2\omega} = \frac 12\left(R^{2\omega}(B) - R^{2\omega}(-B)\right)=\frac12R_0\gamma B\times I$.\cite{ideue_bulk_2017}

The results of our initial measurements are plotted in Fig.~\ref{fig:nonil2ndharm},
carried out before the IL was applied, at \SI{2.5}{\kelvin}.
The insets show the orientations of the magnetic and electric fields.
$\Delta R^{2\omega}$ is measured as a function of $B$,
the value of $\gamma$ is proportional to the slope.

To verify that the measured signal originates from the non-reciprocal resistance of BiTeBr and not some other effect,
we took advantage of the relation from the first section, $\Delta R^{2\omega}\propto B\times I$.
As Fig.~\ref{fig:nonil2ndharm} demonstrates, the signal disappears when the fields are parallel (blue curve),
indicating that the source of the signal is indeed the non-reciprocal resistance of BiTeBr.
Since $\gamma$ is size dependent, $\gamma'=\gamma A$ is introduced to compare different devices,
where $A$ is the cross-section of the sample perpendicular to the current's direction.
In the case of perpendicular $B$ and $E$ vectors (orange curve) $\gamma'=\SI[separate-uncertainty=true]{1.2(3)e-13}{\m\squared\per\tesla\per\A}$,
which is consistent with previous results.\cite{ideue_bulk_2017}
The current and temperature dependence of $\gamma'$ was also measured,
the results are shown in Fig.~S3.
To determine the carrier density $n$,
Hall measurements were carried out (for details see Fig.~S2),
and yielded $n=\SI[separate-uncertainty=true]{1.19(1)e19}{\per\cm\cubed}$,
which is in good agreement with the usual doping of BiTeBr.\cite{ideue_bulk_2017, ogawa_photocontrol_2014}
The relation of $n$ and $\gamma'$ is in good agreement with previous results.\cite{ideue_bulk_2017}

Placing the IL on top of the sample the same second harmonic and Hall measurements were repeated and no change in $\gamma'$ or $n$ were observed,
thanks to the hBN protection layer.
As $\gamma'$ increases with decreasing temperature, all measurements were carried out at \SI{2.5}{\K}.

%%% IL gated meas.

\begin{figure*}[h!]
	\setcitestyle{numbers}
	\includegraphics{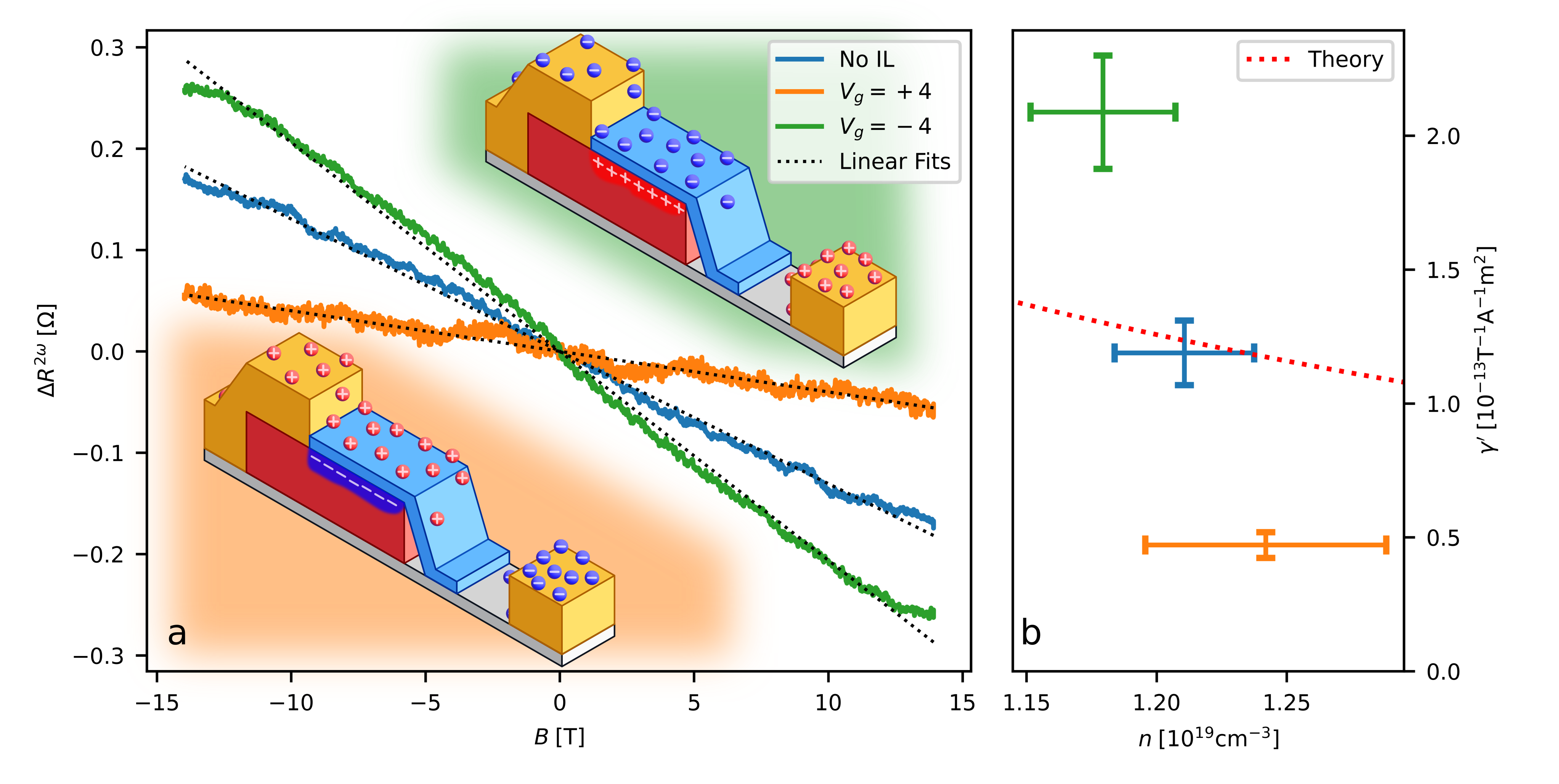}
	\caption{
		\label{fig:gated}
		\emph{a)}~The magnetic field dependence of the second-harmonic signal
		at different gate voltages.
		The insets demonstrate the position of the ions in the IL
		and the induced density changes in the BiTeBr  in the different experiments.
		\emph{b)}~$\gamma'$ as a function of the carrier density extracted from Hall measurements.
		The dashed line shows theory from Ref.~\cite{ideue_bulk_2017}
		without any fitting parameters.
		The divergence from theory can be explained by considering,
		that the second harmonic signal comes mainly from the gated surface layer,
		while the Hall measurements used for determining $n$ measure the whole crystal.
	}
\end{figure*}

In order to probe the density-dependence of $\gamma$, we changed the electron density,
by heating the sample up above the glass transition temperature of the IL, and recooling it with the IL gate set to $V_\mathrm{g}=$\SI{-4}{\volt} first.
After measurements, the process was repeated for \SI{+4}{\volt}.
The results of the second harmonic measurements and densities obtained from Hall measurements are shown in Fig.~\ref{fig:gated},
the full Hall measurements are shown in Fig.~S2.
Comparing the curves on Fig.~\ref{fig:gated} a,
it is easy to see that the slope of the curve,
and therefore $\gamma'$ is enhanced by the gating.
This is consistent with the theory, predicting $\gamma'\propto\frac{1}{n^2}$.

The extracted values for $\gamma'$ as a function of carrier density are plotted in Fig.~\ref{fig:gated}.b.
It is clearly visible that we could tune $\gamma'$ by over \SI{400}{\percent} using IL gating,
with a small modification of less than \SI{10}{\percent} of the total bulk carrier density.

The dashed line in the figure shows the results of calculations based on
the solution of the second order Boltzmann equations from Ref.{\setcitestyle{numbers}\cite{ideue_bulk_2017}}.
Although the trend in the measured values follows the theoretical expectations,
the overall change in $\gamma'$ is much stronger than expected.

The observed stronger dependence can be explained by the charge carriers screening the gate-induced electric field.
Based on the measured electron density of $n=$\SI[]{1e19}{\per\cm\cubed},
the screening length ($r_{\mathrm{TF}}$) was estimated using the Thomas-Fermi approximation\cite{solyom_fundamentals_2010},
with the assumption of a free electron gas.
This yields a value of $r_{\mathrm{TF}}=\SI{2}{\nm}$.
Since this is much shorter than the thickness of the BiTeBr crystal (\SIrange{30}{40}{\nm}),
the electron density is not homogeneous vertically in the Rashba crystal:
the gating induced density change is more pronounced at the top of the sample
and decreases further from the surface as shown in the insets of Fig.~\ref{fig:gated}.a.
This leads to a stronger variation in the rectification effect in the top region of the sample,
without significantly affecting the rest of the crystal,
leading to a higher than expected change in $\gamma'$.
Additionally, BiTeBr is a polar crystal\cite{shevelkov_crystal_1995, fiedler_termination-dependent_2015}
thus the electron density could be significantly different at the surfaces\cite{sakano_strongly_2013, eremeev_rashba_2013, moreschini_bulk_2015}.
The interplay of the strong gating with surface charges and termination-dependent surface states
could also contribute to the observed deviation.

A more homogeneous charge carrier density might be achieved for thinner crystals.
However, usual exfoliation techniques failed to produce flakes thinner than \SI{40}{\nm}.
In one of our recent works\cite{fulop_exfoliation_2018} we have shown that using a special exfoliation method
exploiting gold surface chemistry, it is possible to exfoliate a single layer of BiTeI,
which has the same crystal structure and very similar physical properties as BiTeBr.
With subsequent chemical etching of Au, a single layer of BiTeBr might be fabricated.
In such flakes ionic gating is expected to induce a significantly larger electron density change.
Since $\gamma'$ is proportional to $\frac{1}{n^2}$,
one can expect that the non-reciprocal response can be increased by orders of magnitude,
as the carrier density is reduced further.

In conclusion,
we have demonstrated that thin van der Waals insulator layers allow us to carry out ionic liquid gating experiments
on crystals which are chemically unstable in ionic-liquid environments.
\SIrange{3}{5}{\nm} thin hBN layers serve as perfect protection for sensitive crystals with vertical dimensions as high as
\SIrange{30}{50}{\nm}, with metallic contacts cutting through the hBN insulator.
This novel heterostructure allowed the enhancement of the non-reciprocal response of
the giant Rashba spin-orbit crystal, BiTeBr.
With in situ tuning the non-reciprocal response was changed by over \SI{400}{\percent}.

By exploiting these results, devices with gate tunable bulk rectification effects could be realized,
even in the absence of pn junctions.
Tuning of the Fermi level could also be used for spintronic applications.\cite{kovacs-krausz_electrically_2020}
Giant Rashba materials are ideal candidates for spin-orbit torque devices as well.\cite{manchon_current-induced_2019}
It has been suggested that pressurizing BiTeBr can induce a topological phase transition.\cite{ohmura_pressure-induced_2017}
It is also a promising choice for topological superconductivity when contacted with superconducting electrodes,
due to the large Rashba spin-orbit coupling.\cite{ren_topological_2019, fornieri_evidence_2019, mayer_phase_2019, pientka_topological_2017, hell_two-dimensional_2017}
In such systems, tuning the carrier density could be an important tool to enter into the topological regime.

\textit{ Author contributions} -- BiTeBr crystals were grown by O. E. Tereshchenko and K. A. Kokh
hBN single crystals were grown by T. Taniguchi and K. Watanabe.
Samples were fabricated by M. Kocsis and O. Zheliuk.
AFM characterization was carried out by M. Kocsis.
Raman spectroscopy was carried out by M. Kocsis and P. Kun.
Measurements were carried out by M. Kocsis, O. Zheliuk and E. Tóvári.
P. Makk, Cs. Szabolcs and J. Ye supervised the project.

\textit{ Acknowledgements} --
Authors thank I.~Lukacs, and J.~Ferenc for their help in sample fabrication,
S.~Lenk for their help with AFM measurements,
M. G.~Beckerna, F.~Fülöp, and M.~Hajdu for their technical support,
and T.~Fehér and A.~Virosztek.

This work has received funding from Topograph, CA16218 by COST, 
the Flag-ERA iSpinText project, and from the OTKA FK-123894 and OTKA NN-127900 grants.
M.K. was supported by the ÚNKP-19-3 New National Excellence Program of the Ministry for Innovation and Technology.
P.M. acknowledges support from the Bolyai Fellowship,
the Marie Curie grant and the National Research,
Development and Innovation Fund of Hungary within the Quantum Technology National Excellence Program (Project Nr. 2017-1.2.1-NKP-2017-00001).

O.E.T. and K.A.K. were supported by the Russian Science Foundation (No 17-12-01047).

K.W. and T.T. acknowledge support from the Elemental Strategy Initiative
conducted by the MEXT, Japan ,Grant Number JPMXP0112101001,  JSPS
KAKENHI Grant Number JP20H00354 and the CREST(JPMJCR15F3), JST.

P.K. was supported by the NanoFab2D ERC Starting Grant project.

\textit{ Methods} -- Single crystals of BiTeBr were grown by a modified Bridgman method with rotating heat field\cite{kokh_application_2005}.
Mixtures of binary compounds Bi\textsubscript{2}Te\textsubscript{3} and BiBr\textsubscript{3} were used as charges to grow BiTeBr.
According to Ref.{\setcitestyle{numbers}\cite{petasch_investigations_1999}}, BiTeBr has a congruent melting point at \SI{526}{\celsius}.
Therefore, a stoichiometric charge of the binary compounds was used to grow BiTeBr.
Charges, sealed under vacuum in quartz ampoules, were at first prereacted at temperatures exceeding the melting point by \SI{20}{\celsius}
and then pulled through a vertical gradient of \SI[per-mode=fraction]{15}{\celsius\per\cm} at a rate of \SI{10}{\mm} per day.
More technical details can be found in {\setcitestyle{numbers}Ref.~\cite{kokh_application_2005} and Ref.~\cite{sklyadneva_lattice_2012}}.

BiTeBr was exfoliated onto \SI{290}{\nm} SiO\textsubscript{2} covered wafers using various blue tapes.
Different environments were also experimented with for the exfoliation process.
While the overall number of flakes could be varied, minimal thickness was unaffected.
Due to the thickness of the flakes, optical classification by height was not possible;
AFM measurements were used to select the thinnest flakes.

Classification of the flakes by Raman spectroscopy was essential to verify their composition,
and was carried out for all flakes.
Two different types of impurities were identified in the BiTeBr crystals.
The first resulted in colourful flakes, sometimes as thin as 5 nm.
These were electrically insulating, and were identified as BiOBr based on their Raman and EDS spectra.
The second type was optically indistinguishable from BiTeBr flakes, their thickness was in the same range as well.
This type of crystal did not show the non-reciprocal effects discussed above,
and were identified a Bi\textsubscript{2}Te\textsubscript{3}
based on their Raman\cite{richter_raman_1977, goncalves_optimization_2010, shahil_micro-raman_2012} and EDS spectra.
This led us to believe that the Raman spectra of BiTeBr as published by Sklyadneva \textit{et. al},\cite{sklyadneva_lattice_2012}
are in fact the spectra of Bi\textsubscript{2}Te\textsubscript{3}.

hBN flakes were exfoliated onto a Si/SiO\textsubscript{2} chip,
with an oxide layer thickness of \SI{90}{\nm}.
The flakes were first optically classified using a bandpass filter,\cite{gorbachev_hunting_2011}
and the selected flakes were measured with AFM.
Chosen flakes were transferred onto the BiTeBr flakes,
using a dry transfer technique.\cite{geim_van_2013}
One such structure is shown in Fig.~\ref{fig:sample}.b.
A PMMA layer was applied, and the design of the electrodes was exposed using electron-beam lithography.
The hBN was etched away, using reactive ion etching with a mixture of CF\textsubscript{4} (45 sccm) and O\textsubscript{2} (5 sccm).
This etches the hBN very quickly, without affecting the BiTeBr or the substrate, thus the duration is not critical,
usually \SI{15}{\sec} was chosen.
Ti/Au (\SI{5}{\nm}/\SI{60}{\nm}) electrodes were evaporated onto the sample, using the same mask.
This ensures that the etched windows in the hBN line up perfectly with the electrodes,
sealing the BiTeBr flake away from the IL perfectly.

$n$ could not be measured in the $V_\mathrm{g}=\SI{+4}{\V}$ case as some contacts were damaged,
and the sample became unmeasurable.
The value presented was estimated, by considering the change in $n$ to be the same for both gate voltages.

Measurements were carried out in a PPMS at \SI{2.5}{\kelvin} using low frequency lock-in techniques.
The sample holder was equipped with a rotator which allowed us to carry out both second harmonic (in-plane B)
and Hall-measurements (out-of-plane B) without heating the sample up between measurements.
The IL was applied at room temperature, and set at 220 K.
\bibliography{main,extras}

\end{document}